\documentstyle[preprint,aps]{revtex}
\begin{document}
\draft
\preprint{SI-95-TP3S1}
\title   {Baryon-baryon interactions in large ${\bf N_C}$ \\ chiral
perturbation theory
}

\author{
Bernd Schwesinger\footnote{Electronic address:
schwesinger@hrz.uni-siegen.d400.de}
}

\address{ Siegen University, Fachbereich Physik,
                                  57068 Siegen, Germany }

\maketitle

\begin{abstract}
Interactions of two baryons are considered in large $N_C$ chiral
perturbation theory and compared to the interactions derived from the
Skyrme model. Special attention is given to a torus-like configuration
known to be present in the Skyrme model.
\vspace{2cm}
\end{abstract}

\pacs{PACS numbers  13.75.Cs, 12.39.Dc, 12.39.Fe, 12.40-y}

\narrowtext

\section{Introduction}
\label{sec1}
Chiral perturbation theory relates various low-energy properties of
hadronic systems by means of effective actions. Such effective lagrangians
are defined in terms of the degrees of freedom  manifest in elementary
excitations of the system.
Constraints imposed on the effective action originate from
the symmetry requirements only, as deduced from the observed
particle spectra, from which fact the generality of the method may be
understood\cite{Wnbg79}. In the absence of baryonic degrees of freedom this
calculational scheme has lead to an impressive amount of statements on
properties of mesonic systems, which commonly have been derived and
thus are valid up to fourth chiral order, i.e. to one-loop
level\cite{GL84,GL85}.
Inclusion of the baryons into the
scheme unfortunately introduces complications due to the fact that a
counting scheme based on chiral orders no longer limits the number of
terms appearing at a definite chiral order\cite{GSS88}. Baryons included, the
perturbative expansion
looses much of its usefulness unless further criteria of smallness are
introduced from outside. One such criterion is the order in an
expansion in terms of the number of colors $N_C$ appearing in the underlying
more fundamental theory, quantum chromodynamics\cite{Witten79,Jen93}.

Restricted to purely mesonic systems the effective lagrangians of chiral
perturbation theory are identical to those of the Skyrme model\cite{Sky61}, if
parameters are chosen accordingly: both involve the Goldstone
bosons of spontaneously broken chiral symmetry as principle degrees of
freedom. In the presence of baryons the similarity of the two approaches
seems to disappear, since the lagrangian of chiral
perturbation theory introduces baryons explicitly coupling them to the
Goldstone bosons in $U=\exp {i \tau\cdot \pi /f_\pi}$ via a vertex
\begin{eqnarray}\label{pin-cplng}
{\cal L}_{\pi N}= \frac{1}{2} g_A^{\!\!\!\circ} \bar N \gamma_\mu \gamma_5
\tau_a N
\cdot \frac{i}{2} {\rm tr} \, \tau_a \sqrt{U^\dagger} \partial^\mu U
\sqrt{U^\dagger} \quad
\end{eqnarray}
of order $\sqrt{N_C}$,
whereas the Skyrme model has no such couplings ($g_A^{\!\!\!\circ} = 0$):
baryons in the Skyrme model only appear as topological knots in the meson
fields. More recent developments\cite{Man94,DHM94}, however, are now suggesting
that even in the presence of baryons both approaches, at least to
leading order in $N_C$, are identical in the limit that the extension
of the bare baryon tends to zero. In section 1. I will repeat
this suggestion
thus introducing formalism and notations for the present work, which
otherwise is concerned with the interactions of two baryons.

Systems with baryon number $B = 2$ have had a somewhat peculiar
status in the $SU(2)$ Skyrme model since the minimal energy
configuration is of torus-like structure\cite{KS87,V87} as far as
energy and baryon
number density are concerned thus displaying only a very remote
resemblance to two interacting baryons. The natural question to ask then
is: if large $N_C$ chiral perturbation theory leads to the Skyrme model
for systems containing one baryon, does the interaction of two baryons
lead to torus-like structures in chiral perturbation theory? Since the
foundations of chiral perturbation theory are firmly
established, the answer to the question posed is of principle
importance. I will attempt to answer the question in two steps.
In section 2. I first will show, that the interaction between two
baryons in large $N_C$ chiral perturbation theory at large separations
is identical to the Skyrme model expressions. In section 3. I will
examine the short distance behaviour, which is only accessible
numerically, displaying the results as to make clear, that large $N_C$
chiral perturbation theory will indeed lead to torus-like
configurations when two baryons are approaching one another
adiabatically. Latter assumption is, of course, inherent at leading
order in $N_C$.

\section{Large $N_C$ chiral perturbation theory and the skyrmion}
\label{sec2}
The gap between chiral perturbation theory and the Skyrme
model is bridged by the observation\cite{GS84,DM93} that the
$\pi N$-scattering amplitude, which is of order one in $N_C$-counting,
can only emerge once all order $N_C$ diagrams,
which are present due to the  coupling of the baryonic
axial current to the mesonic one (order $\sqrt{N_C}$) in
(\ref{pin-cplng}), have cancelled.
Without such cancellations $\pi N$-scattering would be of order $N_C$.
The cancellation requires an infinite tower of baryonic states, all
degenerate at order $N_C$, having their spins equal to their isospins.
The generalization of the $\pi N$-coupling (\ref{pin-cplng})
to the whole tower is obtained by substitution
\begin{eqnarray}
\gamma_i \gamma_5 \tau_a \rightarrow X^{a\,i} \nonumber
\end{eqnarray}
of the spin-isospin matrixelements for $I=J=\frac{1}{2}$ representations,
by those for the whole tower. The couplings which lead to a
cancellation of the order $N_C$ scattering amplitude have been
determined in ref.\cite{GS84,DM93}:
\begin{eqnarray}
\langle I'=J',I'_3,J'_3 \mid X^{\alpha \beta}
\mid I=J,I_3,J_3 \rangle =
\qquad \qquad\qquad\qquad \qquad\qquad\qquad \qquad\qquad \\
-\sqrt{(2I'+1)(2I+1)}
(-)^{I'-I_3'} \left(
\begin{array}{ccc}
I'    &   1    & I    \\
-I'_3 & \alpha & I_3
\end{array} \right)
(-)^{J'-J_3'} \left(
\begin{array}{ccc}
J'    &   1    & J    \\
-J'_3 & \beta & J_3
\end{array} \right) =
 \nonumber \\
\int d[A] \,
\sqrt{\frac{2I'+1}{8 \pi^2}}(-)^{I'-I'_3}D^{* \,I'=J'}_{-I'_3J'_3}(A) \,
(-)^{\alpha} D_{-\alpha \beta}(A) \,
\sqrt{\frac{2I+1}{8 \pi^2}}(-)^{I-I_3}D^{I=J}_{-I_3J_3}(A) .\nonumber
\end{eqnarray}
Note, that spherical indices have been used here. The second part of
the equation
is just an identity for the D-functions of matrices $A\in
SU(2)$. This identity will prove to be
useful, once we have rotated the infinite tower of degenerate baryon states
$\mid I=J,I_3,J_3\rangle $
to a basis classified according to the orientations $ A $ of a
baryon in isospace:
\begin{eqnarray}\label{basis}
\mid A \rangle = \sum_{I,I_3,J_3}
\sqrt{\frac{2I+1}{8 \pi^2}}(-)^{I-I_3}\,D^{* \, I=J}_{-I_3J_3}(A)\,
\mid I=J,I_3,J_3\rangle .
\end{eqnarray}
The new basis of degenerate baryons diagonalizes the pion-baryon
coupling
\begin{eqnarray}\label{xia}
X^{a\,i} = \int d[A] \,  D_{a i}(A)\,
  \mid A\rangle \langle A \mid
\end{eqnarray}
as may easily be seen by insertion of (\ref{basis}) into (\ref{xia}).
Thus, in leading order in $N_C$ baryons do not change their orientation
in isospace when interacting with pions. For similar reasons baryons
do not move in space upon interaction with the mesons since their
velocities are of order $1/N_C$: baryons behave like a static source of
fixed orientation $A$ and position ${\bf X}$ for the pion fields.
The large $N_C$
interactions of pions and baryons are then summarized by the following
lagrangian, the leading $N_C$-dependence of which has been factored to the
front of the lagrangian:
\begin{eqnarray}\label{lagrangian}
{\cal L}&=&N_C \left[ {\cal L}^{({\rm meson})}(U) +
{\cal L}^{({\rm source })}(U;A;{\bf X}) \right] \\
{\cal L}^{({\rm meson})}&=&\frac{f_\pi^2}{4 N_C} {\rm tr} \,
\partial_\mu U \partial^\mu U^\dagger
+\frac{f_\pi^2 m_\pi^2}{4 N_C} {\rm tr} \,( U +U^\dagger -2)
+ \cdots \nonumber \\
{\cal L}^{({\rm source })}&=& -\frac{3}{2 N_C} g_A^{\!\!\!\circ} \,
\Delta ({\bf x} - {\bf X}) D_{ai}(A) \,
\cdot \frac{i}{2} {\rm tr} \, \tau_a \sqrt{U^\dagger} \partial_i U
\sqrt{U^\dagger}  .\nonumber
\end{eqnarray}
${\cal L}^{({\rm source })}$ and ${\cal L}^{({\rm meson})}$ are of order
$(N_C)^0$.
For spin=isospin=$\frac{1}{2}$ states the matrix elements of the
$D$-function in (\ref{lagrangian}) are given by  $D_{ai}
\rightarrow -\frac{1}{3} \tau_a \sigma_i$, so the $\pi N$-coupling
implicit in (\ref{lagrangian}) coincides with the one given earlier in
(\ref{pin-cplng}).

In a very readable recent publication Manohar\cite{Man94} demonstrates
that an $N_C$-independent regularization of the functional integral
constraining the effective lagrangian(\ref{lagrangian}) to its range of
validity, i.e. to small momentum scales, can be achieved by giving the
baryon source a finite extension $R_0 \sim 1 {\rm GeV}^{-1}$
\begin{eqnarray}\label{source}
\Delta ({\bf x} - {\bf X})=\bigl( \frac{4 \pi}{3}R_0^3 \bigr)^{-1}
\, \Theta(R_0-\mid {\bf x}-{\bf X} \mid ).
\end{eqnarray}
In this case a factor $N_C / \hbar $ multiplies the exponent in the
integrand of the functional integral for which the leading terms in an
expansion in powers of $1/N_C$ therefore turn out to be
equivalent to the leading terms in a semiclassical expansion in powers
of $\hbar$.
Thus, the leading terms in $N_C$ of the pion-baryon interactions are
obtained by solution of classical equations of motion for the pion cloud
around a static baryon source of fixed position and isospin orientation!

The structure of the pion cloud around such a fixed source which
satisfies the classical Euler-Lagrange equations has the form
\begin{eqnarray}\label{ansatz}
U= A\, {\rm e}^{i \tau \cdot \hat x \, \chi (x)}\, A^\dagger
\end{eqnarray}
and the cloud is completely determined by solution of a second
order radial differential equation for the remaining chiral angle $\chi$
\begin{eqnarray}\label{eom}
\partial_x ( x^2 \partial_x \chi) - \sin 2\chi - m_\pi^2 x^2 \sin \chi
+ \cdots
= \frac{3 g_A^{\!\!\!\circ}}{2 f_\pi^2} \left[ \partial_x \Delta(x)
-\frac{2}{x}(1-\cos \chi) \Delta(x) \right].
\end{eqnarray}
The dots stand for higher order terms from ${\cal L}^{({\rm meson})}$
in (\ref{lagrangian}).

In figure 1. we display the chiral angle of the cloud as a function of
distance from the center of the baryon source. Coming from large
distances the chiral angle has the one-pion tail
\begin{eqnarray}\label{chiasy}
\chi(x) \stackrel{x \rightarrow \infty}{\longrightarrow}
\frac{3 \tilde g_A}{8 \pi f_\pi^2} m_\pi^2
\bigl(1+\frac{1}{m_\pi x}\bigr)
\frac{1}{m_\pi x} {\rm e}^{-m_\pi x}
\end{eqnarray}
where $\tilde g_A=f_\pi/M_N \, g_{\pi NN}$ is the physical $\pi
N$-coupling. $\tilde g_A$ differs from the axial charge $g_A$ by terms
of order ${\cal O}(m_\pi^2)$. At $x=R_0$ the source
enforces a discontinuity in the derivatives proportional to the
bare $\pi N$-coupling $g_A^{\!\!\!\circ}$. The discontinuity is adjusted, of
course, by making the total solution regular at the origin.

For a large source, i.e. a small cutoff, the lowest order terms (w.r.t.
$\chi$) of the
lagrangian (\ref{lagrangian}) are sufficient. The physical $\pi
N$-coupling equals
\begin{eqnarray}\label{ga}
\tilde g_A=-3 \left\{ \frac{{\rm sinh}(m_\pi R_0)}{(m_\pi R_0)^3}-
\frac{{\rm cosh}(m_\pi R_0)}{(m_\pi R_0)^2} \right\} g_A^{\!\!\!\circ}
= \bigl( 1+\frac{1}{10} (m_\pi R_0)^2+\,{\cal O}(m_\pi R_0)^4 \bigr)
g_A^{\!\!\!\circ}
\end{eqnarray}
and the mass shift
$\delta M$ of the baryon due to the cloud may also be calculated
analytically\cite{Man94} as
\begin{eqnarray}\label{mass1}
\delta M &=& \frac{81 {g_A^{\!\!\!\circ}}^2}{64 \pi f_\pi^2 m_\pi^3 R_0^6}
(1+m_\pi R_0)\left[(1-m_\pi R_0)-(1+m_\pi R_0){\rm e}^{-2 m_\pi
R_0}\right] \\
&=&\qquad
\quad-\quad\frac{27 {g_A^{\!\!\!\circ}}^2}{32 \pi f_\pi^2 R_0^3}
\quad+\quad\frac{27 {g_A^{\!\!\!\circ}}^2 m_\pi^2}{80 \pi f_\pi^2 R_0}
\quad-\quad\frac{9 {g_A^{\!\!\!\circ}}^2 m_\pi^3}{32 \pi f_\pi^2 }
\quad+\quad{\cal O}(R_0) \nonumber
\end{eqnarray}
The first two (for $R_0 \rightarrow 0$ singular) cutoff-dependent shifts
may be absorbed into the constants of the bare lagrangian of chiral
perturbation theory: term
one into the chiral invariant baryon mass term, term two into the quark
mass contribution to the nucleon mass which is
proportional to $m_q \sim m_\pi^2$. The third term is
non-analytic in the quark masses and cannot be reabsorbed into the bare
lagrangian, where no such terms are present: the third term,
independent of the cut-off, is a genuine
finite correction and identical to the one-loop correction
to the baryon mass as calculated in standard chiral perturbation
theory\cite{GSS88,Jen92}
with intermediate nucleon and isobar states.

Due to the multivaluedness of the lagrangian,
the requirement of regularity of the energy density only
demands the chiral angle to be some multiple of $\pi$ at the origin.
Returning to figure 1. we see, that the bare pion-baryon
coupling $g_A^{\!\!\!\circ} \rightarrow 0$ if the chiral angle, coming
from large
distances where it is fixed, just reaches a multiple of $\pi$ at the
origin. In such a case we have a finite renormalized pion-baryon
coupling $\tilde g_A$ in a purely mesonic theory since the bare
pion-baryon coupling
$g_A^{\!\!\!\circ}$ now is zero: this configuration of the cloud is
identical to the chiral field of the skyrmion\cite{DHM94}.

\section{Asymptotic interactions}
\label{sec3}
The presence of two baryons in large $N_C$ chiral perturbation theory
is realized by placing two baryonic sources, one at ${\bf X}/2$ with
orientation $A$ the other at $-{\bf X}/2$ with orientation $B$. As long
as the separation of the two sources $X$ is greater than twice the
radius $R_0$ of the source the interaction proceeds through meson
exchange only, given by a trivial generalization of the lagrangian
(\ref{lagrangian}) for the sources:
\begin{eqnarray}\label{lsource}
{\cal L}&=&N_C \left[ {\cal L}^{({\rm meson})}(U) +
{\cal L}^{({\rm sources })}(U;A,B;{\bf X}) \right] \\
{\cal L}^{({\rm sources })}&=& -\frac{3}{2 N_C} g_A^{\!\!\!\circ} \,
\Delta ({\bf x} -\frac{1}{2} {\bf X}) D_{ai}(A) \,
\cdot \frac{i}{2} {\rm tr} \, \tau_a \sqrt{U^\dagger} \partial_i U
\sqrt{U^\dagger}  \nonumber \\
& & -\frac{3}{2 N_C} g_A^{\!\!\!\circ} \,
\Delta ({\bf x} +\frac{1}{2} {\bf X}) D_{ai}(B) \,
\cdot \frac{i}{2} {\rm tr} \, \tau_a \sqrt{U^\dagger} \partial_i U
\sqrt{U^\dagger}
.\nonumber
\end{eqnarray}
{}From the lagrangian (\ref{lsource}) we may deduce the classical
Euler-Lagrange equations in order to calculate cloud effects to leading
order in $N_C$.

The restriction of the equations of motion to the case of
large sources simplifies matters appreciably.
Then it is sufficient to keep
terms which are maximally linear in the chiral angles $\chi_b$ parametrizing
the matrix $U$. In this case the two sources only appear as inhomogeneous terms
in the equations of motion, independent of the chiral angles. The
solution to such a linear inhomogeneous differential equation is,
of course, a superposition of the chiral fields for each of the
sources separately, as they have emerged from eq.(\ref{eom}) (in its linearized
form):
\begin{eqnarray}\label{chi2}
\chi_b=D_{bi}(A) \, \hat{x}_{-_i} \, \chi_- +
       D_{bi}(B) \, \hat{x}_{+_i} \, \chi_+
\end{eqnarray}
where
\begin{eqnarray}
{\bf x}_-={\bf x}-\frac{1}{2}{\bf X}, \qquad
{\bf x}_+={\bf x}+\frac{1}{2}{\bf X}, \\
\chi_-=\chi(\mid {\bf x}_-\mid), \qquad
\chi_+=\chi(\mid {\bf x}_+\mid) . \nonumber
\end{eqnarray}
The mass shift of the two baryons may be deduced from
\begin{eqnarray}\label{mass2}
\delta M_{B=2} &=&\frac{1}{2} \int d^3x \left\{
-\frac{3}{2}g_A^{\!\!\!\circ} \,
\Delta ({\bf x}_-) D_{bi}(A) \,
\partial_i \chi_b
-\frac{3}{2}g_A^{\!\!\!\circ} \,
\Delta ({\bf x}_+) D_{bi}(B) \,
\partial_i \chi_b \right\}
\end{eqnarray}
where the equations of motion for $\chi_b$ have been used to eliminate
the contributions from the purely mesonic parts of the energy density.
Inserting the chiral fields given in (\ref{chi2}) the mass shift in
(\ref{mass2}) contains the mass shifts of the individual baryons given
in (\ref{mass1}) and an interaction term
\begin{eqnarray}\label{intasy}
V_{{\rm asy}} &=&-\frac{3}{4}g_A^{\!\!\!\circ} \, \int d^3x \left\{
\Delta ({\bf x}_-) D_{bi}(A) D_{bj}(B)
\partial_i  \hat{x}_{+_j} \, \chi_+
+
\Delta ({\bf x}_+) D_{bi}(B) D_{bj}(A)
\partial_i \hat{x}_{-_j} \, \chi_-
\right\}\\
&=&\quad
\frac{3}{4}g_A^{\!\!\!\circ} \, \int d^3x \left\{
 D_{ij}(A^\dagger B)
  \hat{x}_{+_j} \, \chi_+\, \partial_i \Delta ({\bf x}_-)
+
 D_{ij}(B^\dagger A)
 \hat{x}_{-_j} \, \chi_- \, \partial_i \Delta ({\bf x}_+)
\right\} \nonumber \\
&=&\quad
\frac{3}{4}g_A^{\!\!\!\circ} \, D_{ij}(A^\dagger B)\,  \int d^3x \left\{
  \hat{x}_{+_j} \, \chi_+\, \partial_i \Delta ({\bf x}_-)
+
 \hat{x}_{-_i} \, \chi_- \, \partial_j \Delta ({\bf x}_+)
\right\}, \nonumber
\end{eqnarray}
For the last step I have used the fact that the $D$-functions w.r.t. Cartesian
indices are real.
Eq.(\ref{intasy}) is valid for large sources and consequently large
separations $ X $ between the sources. In the integrand each
source multiplies the chiral field of the other source so the
asymptotic form of the chiral angles from (\ref{chiasy}), (\ref{ga})
may safely be inserted leading to the final result
\begin{eqnarray}\label{intasy2}
V_{{\rm asy}} &=&\frac{9}{16 \pi f_\pi^2 }\tilde g_A^2 \,
D_{bi}(A) D_{bj}(B)\, \partial_i \partial_j
\frac{1}{X}{\rm e}^{-m_\pi X}\, \bigl(1 + {\cal O}(m_\pi R_0) \bigr) ,
\end{eqnarray}
where the derivatives now act on ${\bf X}$.

The asymptotic interaction
behaves smoothly as the cutoff is removed and then
precisely equals the expression for the asymptotic interaction of two
skyrmions derived by Skyrme\cite{Sky61} thirty years
ago. Taking its matrixelements for baryons $A$ and $B$, both with
spin=isospin=$\frac{1}{2}$, yields the well known one-pion
exchange potential for two nucleons, because then
$D_{bi}(A) D_{bj}(B) \rightarrow \frac{1}{9}\,
{\mbox{\boldmath $\tau$}}^A \cdot
{\mbox{\boldmath $\tau$}}^B \, \sigma_i^A \, \sigma_j^B$.

\section{Short range interactions}
\label{sec4}
The exploration of the short range behaviour of baryon interactions at
leading order in $N_C$ introduces several speculative elements with
respect to the precise form of the effective action and several
uncertainties in precision, because the investigation has to be
performed numerically, as I will explain. Nevertheless, I believe that
the main ingredients and the main conclusions are under control.

Since we wish to calculate cloud energies at small separations $X$
of two baryons we must ensure that the sources do not overlap,
i.e. the radius of the source must obey $R_0 < \frac{1}{2} X$.
Therefore the volume of the source is small and due to its
normalization to unit baryon charge, eq.(\ref{source}), this will lead
to strong meson fields close to the source. Clearly, the situation can
no longer be handled using the linearized classical Euler-Lagrange
equations and higher order terms in the chiral angles $\chi_b$ are
required. Then, of course, the chirally lowest order terms quoted
explicitly in the lagrangian (\ref{lagrangian}) are no longer
sufficient either.

For the purpose of the present investigation we add
one further fourth order term to the mesonic lagrangian, namely the
fourth order stabilizing term of the Skyrme model:
\begin{eqnarray}
{\cal L}_{{\rm Skyrme}}=\frac{1}{32e^2} \;{\rm tr}\;
[U^\dagger \partial_{\mu} U, U^\dagger \partial_{\nu} U]
[U^\dagger \partial^{\mu} U, U^\dagger \partial^{\nu} U].
\end{eqnarray}
It naturally appears as the larger of the two chirally symmetric terms
in next (i.e. fourth) order chiral perturbation theory\cite{GL84}.

The truncation
to fourth order, which we will apply - also for simplicity -, is a
prejudice.  Nevertheless, it is
motivated by the experience, that other higher order terms in the
Skyrme model do not change the details of the meson cloud beyond say
.25fm and that the Skyrme term is phenomenologically - almost -
sufficient\cite{e4}.

In order to make a meaningful comparison between the $B=2$ sectors of
the Skyrme model and large $N_C$ chiral perturbation theory, we choose
$e=4$ for the Skyrme parameter, which together with $f_\pi=93$MeV
yields a good phenomenological description of baryon and baryon-meson
systems in the former\cite{e4}. For the latter, we fix the bare $\pi
N$-coupling
to $g_A^{\!\!\!\circ}=1.72$ in which case the cloud of the
skyrmion is identical to the cloud around a sharp baryon source of
radius $R_0=.25$fm. Due to numerical problems explained later, we will
actually use a smoother source
\begin{eqnarray}\label{source2}
\Delta ({\bf x})=\bigl( \frac{4 \pi}{3}R_0^3 \bigr)^{-1}
\, \bigl(1+(\frac{x}{R_0})^{20}\bigr)^{-1}.
\end{eqnarray}
for which $g_A^{\!\!\!\circ}=2.01$ will make the meson cloud agree
with the one around the skyrmion in the outside region. In figure 1. I have
displayed the three chiral angles for the cases skyrmion, sharp source,
smooth source as calculated with the parameters quoted here.

Three separations, $X=$.7fm, 1.4fm and 2.1fm, will be considered for
the two sources. Placed at the smallest separation, two skyrmions with
a relative isospin orientation of $A^\dagger B = i \tau_2$
easily deform to a torus-like configuration, as has been shown by
numerical minimizations\cite{Stern89,WW92,CSB92}
on finite three dimensional lattices.
An essential point of
these numerical calculations is that the transition from two solitons
separated along say the $z$-axis leads to a torus with a symmetry axis
perpendicular to the $z$-axis such that axial symmetry cannot be
maintained all time during the transition. Therefore, the numerical
minimization of such configurations requires a general three dimensional
lattice. One immediate consequence is that in three dimensions the
lattice cells will be rather coarse, if one wishes to keep the
computational effort in reasonable limits. Hence, a sharp source is
problematic on a
mesh with a rather low point density and has motivated its substitution
by the smoother counterpart in (\ref{source2}).

The next obstacle one is confronted with in the numerical minimization
comes from parametrizations using three chiral angles, where the
multivaluedness of the angular functions quickly leads to numerical
instabilities on a finite three dimensional mesh. To overcome the
problem, I have switched to a non-unitary parametrization of the chiral
fields
\begin{eqnarray}
U=\Phi_0 + i\,{\mbox{\boldmath $\tau$}}\cdot{\mbox{\boldmath $\Phi$}},
\end{eqnarray}
where unitarity is enforced by a constraint
\begin{eqnarray}
C=\int d^3x \, \lambda({\bf x}) \,
\bigl(\Phi_0^2 + {\mbox{\boldmath $\Phi$}}^2 -1\bigr)^2
\end{eqnarray}
on the four functions.

The non-unique non-unitary extension of the energy functional in terms
of these four functions was chosen as
\begin{eqnarray}
M^{({\rm meson })} = \int d^3x
\left\{ \frac{1}{2}f_\pi^2 \Lambda_i^a \Lambda_i^a +
\frac{1}{4 e^2} \epsilon_{abc}\epsilon_{ade}
\Lambda_i^b \Lambda_j^c \Lambda_i^d \Lambda_j^e +
f_\pi^2 m_\pi^2 (1-\Phi_0) \right\}
\end{eqnarray}
for the purely mesonic parts where the abbreviation
\begin{eqnarray}
\Lambda_i^a = \Phi_0 \partial_i \Phi_a - \Phi_a \partial_i \Phi_0
+ \epsilon_{abc} \Phi_b \partial_i \Phi_c
\end{eqnarray}
has been used. The source terms require a non-unitary extension of the
the square root of $U$. I have used the following form which in
contrast to other possibilities is
numerically non-singular when $\Phi_0 \rightarrow -1$:
\begin{eqnarray}
M^{({\rm sources })}(A,B,{\bf X}) &=&
-\frac{3}{2} g_A^{\!\!\!\circ} \,
\int d^3x
\left\{ \Delta ({\bf x} -\frac{1}{2} {\bf X}) \delta_{ab}
+\Delta ({\bf x} +\frac{1}{2} {\bf X}) D_{ab}(A^\dagger B) \right\} \times
\\& & \qquad \qquad \qquad
\left\{ \Phi_0 \partial_b \Phi_a - \Phi_a \partial_b \Phi_0 +
\bigl( \sqrt{\Phi_0^2 + {\mbox{\boldmath $\Phi$}}^2} - \Phi_0 \bigr)
\Phi \partial_b \hat \Phi_a  \right\} . \nonumber
\end{eqnarray}
$\hat \Phi$ is the unit vector of the fields ${\mbox{\boldmath
$\Phi$}}$ and $\Phi = \mid  {\mbox{\boldmath
$\Phi$}} \mid$. A global isospin rotation $A$ has been performed on the
whole $B=2$ configuration.
Due to this global rotation  $A^\dagger B$ appears as  relative isospin
orientation between the two sources in the functional.
Of course, because of isospin symmetry the global rotation does not affect
the energy density,  $ M^{({\rm sources })}(A,B,{\bf X})=
M^{({\rm sources })}(1,A^\dagger B , {\bf X})$.

Once the unitarity constraints are satisfied exactly,
different extensions of the energy functional would, of course, yield
identical answers. However, since the constraints are only obeyed
approximately in a numerical minimization on a finite mesh, different
extensions lead to differing numbers.

The minimization, finally, varies the four functions
$(\Phi_0 , {\mbox{\boldmath $\Phi$}})$ at every mesh
point independently lowering their contribution to the sum
$M^{({\rm meson })}+M^{({\rm sources })}+C $ for some fixed large
non-negative function $\lambda$ till no further decrease in this sum
occurs. The sum without the contribution of the constraint,
$M^{({\rm meson })}+M^{({\rm sources })} $, is then interpreted as the
minimal energy of the configuration.

The difference between Skyrme model and large $N_C$ chiral
perturbation theory has been reduced to the magnitude of the bare
pion-baryon coupling and the boundary conditions on the meson cloud,
here. Thus, for both cases numerical minimizations may be performed
using the same program and the same three dimensional lattice which is
an advantage in a direct comparison of the two.

A lattice with randomly distributed points, the density of which is
roughly proportional to the expected energy density, has been used
here. It proved to be
superior in precision and stability relative to an equidistant one,
once the same number of points is involved. The price to be paid for
such an advantage is that the energy of a given configuration depends on
the position of where it has been placed on the lattice. I have tested
this dependence for the case of a single smooth source. Its exact
energy as determined from the solution of the differential
equations of motion is $M=-7568$MeV (using the parameters quoted
already). Putting this configuration on the lattice at positions
where the two sources will be located later overestimates the energy by
8\% at $\frac{1}{2} X =$ .35fm, 15\% at $\frac{1}{2} X =$ .7fm, and
20\% at $\frac{1}{2} X =$ 1.05fm. The reason is understood from the
errors in derivatives calculated from finite differences in regions
where the source changes rapidly: further away from the origin of the
lattice the density of points drops.

A numerical calculation of the
interaction energies of two sources as a function of their separation
only makes sense, if the result is compared to the sum of energies of single
sources located at identical positions on the same lattice.
The resulting difference is small
and I estimate its errors to be much better than the $\sim$10\%
deviations between exact and numerically determined absolute mass
shifts. As may be seen from figure 2. the interaction energy determined
this way shows a remarkably smooth dependence on the separation and
actually approaches the analytically determined asymptotic interaction,
also included in the figure.

Since the topological
configuration has no sources, the position dependence of a given
configuration is  much smaller than for the case of explicit sources:
for the $B=1$ soliton
we find $M_{{\rm numerical}}=1752$MeV relative to the exact result
$M=1756$MeV and for the torus configuration in the $B=2$ sector we have
$M_{{\rm numerical}}=3359$MeV relative to the exact result:
$M=3371$MeV. Note, that the huge discrepancy in absolute masses between
topological configurations and the one with explicit sources is
irrelevent, since the latter still miss
the unspecifed bare mass of the source to be added. Let me also emphasize,
that the large soliton mass in the Skyrme model is of
no concern, since the Casimir energies of the soliton appearing in next
to leading order in $N_C$ yield the desired corrections\cite{M93} (at
least for the parameters used here).

In figure 2. I have only displayed the interaction energy of two sources
at a fixed relative orientation $A^\dagger B = i\tau_2$ which is the
most attractive channel for skyrmion-skyrmion interactions. For the
smallest separation $X=.7$fm, where the Skyrme model finds maximal
attraction in a torus-like configuration, I have tested the isospin
dependence of the interaction between explicit sources for two other
cases $A^\dagger B = i\tau_3$ and $A^\dagger B = 1$: both orientations
lead to a repulsive interaction  of +780MeV in the first and
+625MeV in the second case indicating that the most attractive
orientation $A^\dagger B = i\tau_2$ with -290MeV interaction
energy for the explicit sources
is identical to the Skyrme model case. So there is a qualitative
agreement between both, but quantitatively differences are rather large.
Unfortunately, I have no possibility to check, whether the quantitative
differences depend on the arbitrary extension of the source, since I
cannot make it larger at separation $X=$.7fm without having the sources
overlapping and I cannot make it smaller either, because the numerical
problems become unmanageable.

There is, however, indirect evidence that the extension of the source
plays a major role quantitatively: coming to the central point of the
present investigation we now compare the minimal energy density of two sources
separated by .7fm with a relative isospin orientation of $A^\dagger B =
i\tau_2$ to the energy density of the Skyrme model's torus.

This
comparison is presented in the figures 3a,b - 5a,b which display these
densities in three orthogonal planes with the origin as common
point. Figures 3a,b show the plane orthogonal to the $y$-axis, figure
3a for explicit sources, figure 3b for the torus, which evidentally
has thus been cut perpendicular to its symmetry axis. Figure 4a,b show
the corresponding cuts orthogonal to the $x$-axis and in figure 4b
one sees the two
bumps where the torus has been cut parallel to its symmetry axis.
Figures 5a,b finally show the cut perpendicular to the $z$-axis. Due to
the axial symmetry of the torus, this cut leads to an identical density
distribution as the one in figure 4b. Since the explicit sources are
separated in $z$-direction, the plane in figure 5a does not cut the
sources and one only sees the positive definite energy density of the
meson cloud arranged in a way very similar to the torus, albeit lower. Closer
inspection of figure 4a, where now the sources have been cut, inside of
which the energy density is high and negative, one may realize that the
meson cloud outside is again very similar to the case of the torus,
figure 4b: low density between the sources leading to two isolated
bumps, were it not for the holes punched into them by the sources.

Returning to figure 3a now we are no longer surprised to find, that
outside of the holes made by the sources the meson cloud has arranged
itself in form of a ring with an additional low positive density hole
in the center, just as in figure 3b. It only appears, that the ring
formed by the meson cloud is slightly deformed by the presence of the
source, since the latter has a finite extension. I am confident, that
sources of smaller extensions will lead to energy densities which will
come even closer to the torus configuration, so I suspect that
quantitative differences in the interaction energies are mainly due to
the finite extension of the source.

There remains one
interesting question unanswered: in contrast to the case with
topological solitons, the distance between the explicit sources is a
well defined quantity, so one can ask what happens, when the two
sources approach each other even closer than the separation at
which the torus forms from two initial $B=1$ solitons? As emphasized
already, the answer is, unfortunately, beyond the numerical abilities of the
calculation outlined here.

\section{Conclusions}
\label{sec5}
The present investigation has been dealing with the interactions of two
baryons in chiral perturbation theory at leading order in $N_C$. Such an
investigation has become feasible due to an observation by
Manohar\cite{Man94} that the leading order interactions may be obtained by
solution of classical Euler-Lagrange equations. These describe the pion fields
around  static baryon sources fixed at a definite position in space with a
definite orientation in isospace. The situation is clearly reminiscent
of the Skyrme model to the results of which we have made direct comparison.

Firm statements can be made for the long-range interaction, because in
this case there are no uncertainties in the chirally lowest order terms
of the effective action which are sufficient here. The long-range
interaction turns out to be identical to the
long-known one-pion exchange interaction, a result that
certainly is not unexpected. Furthermore, the long-range force is
identical to the
interaction derived from the Skyrme model, so in one more respect this
suggests that large $N_C$ chiral perturbation theory and the Skyrme
model are actually the same language.

If true, one must worry about a peculiar
field configuration known in the $B=2$ sector of the Skyrme model,
namely a torus-like configuration of the meson cloud which represents
the classical energy minimum located at small separations.
Its reception among intermediate
energy physicists has been ambivalent, ranging from 'looking at it as
an artifact of the model' to 'accepting it as the origin of attraction
between nucleons'.

I have tried to explore the case of the torus in the framework
of large $N_C$ chiral perturbation theory, but in doing so, I had to
add speculative elements to the investigation as far as chirally
higher order terms of the effective action were concerned. Specifically,
a simplifying assumption had to be made, that the main term of fourth chiral
order, the well-known Skyrme stabilizer is sufficient to describe the
physics of the meson cloud down to distances of a quarter of a fermi.
Although this cannot be entirely correct quantitatively,
corrections from other higher order terms will certainly not upset the
outcome of this investigation, which is: a torus-like meson cloud
also appears around explicit bare baryon sources in leading order $N_C$ just
as in the Skyrme model.
The configuration will be stable with respect
to modifications in the effective action, because the torus in the Skyrme
model has been stable against such changes.

\section{Acknowledgements}
\label{sec6}

I gratefully acknowledge the help of V. B. Kopeliovich and B. Stern
who allowed me to upgrade and
modify their SU(3) minimization program for the present investigation.
I thank the Institute for Nuclear Theory at the University of
Washington for its hospitality and the Department of Energy as well as
the Deutsche Forschungsgemeinschaft for partial support during the
completion of this work.

\newpage

\narrowtext

\begin{figure}
\caption{The chiral angles of the meson cloud around a baryon source with
sharp cut-off (full line), with smooth cut-off (dashed line) and for a
topological configuration (thin full line).}
\label{fig1}
\end{figure}

\begin{figure}
\caption{Energies of various baryon number $B=2$ configurations as a
function of their separation. The full line displays the interaction of
two topological solitons in the product ansatz, which asymptotically
equals the one-pion exchange force in equation (17).
The dot
marks the position of the topological torus configuration, the distance
of which is defined by the separation in the product ansatz, which
after minimization deforms to the torus. The vertical bars
show the energy including an error estimate for
two smooth explicit baryon sources with relative isospin orientation
$A^\dagger B = i \tau_2$.}
\label{fig2}
\end{figure}

\begin{figure}
\caption{Energy density in the plane containing the origin
and perpendicular to the $y$-axis for  (a) two smooth explicit baryon
sources with
relative isospin orientation $A^\dagger B = i \tau_2$ at .7fm
separation along the $z$-axis,
(b) the topological torus configuration.}
\label{fig3}
\end{figure}

\begin{figure}
\caption{Energy density in the plane containing the origin
and perpendicular to the $x$-axis for (a) two smooth explicit
baryon sources with
relative isospin orientation  $A^\dagger B = i \tau_2$ at
.7fm separation along the $z$-axis,
(b) the topological torus configuration.}
\label{fig4}
\end{figure}

\begin{figure}
\caption{Energy density in the plane containing the origin
and perpendicular to the $z$-axis for (a) two smooth explicit
baryon sources with
relative isospin orientation $A^\dagger B = i \tau_2$
at .7fm separation along the $z$-axis,
(b) the topological torus configuration.}
\label{fig5}
\end{figure}


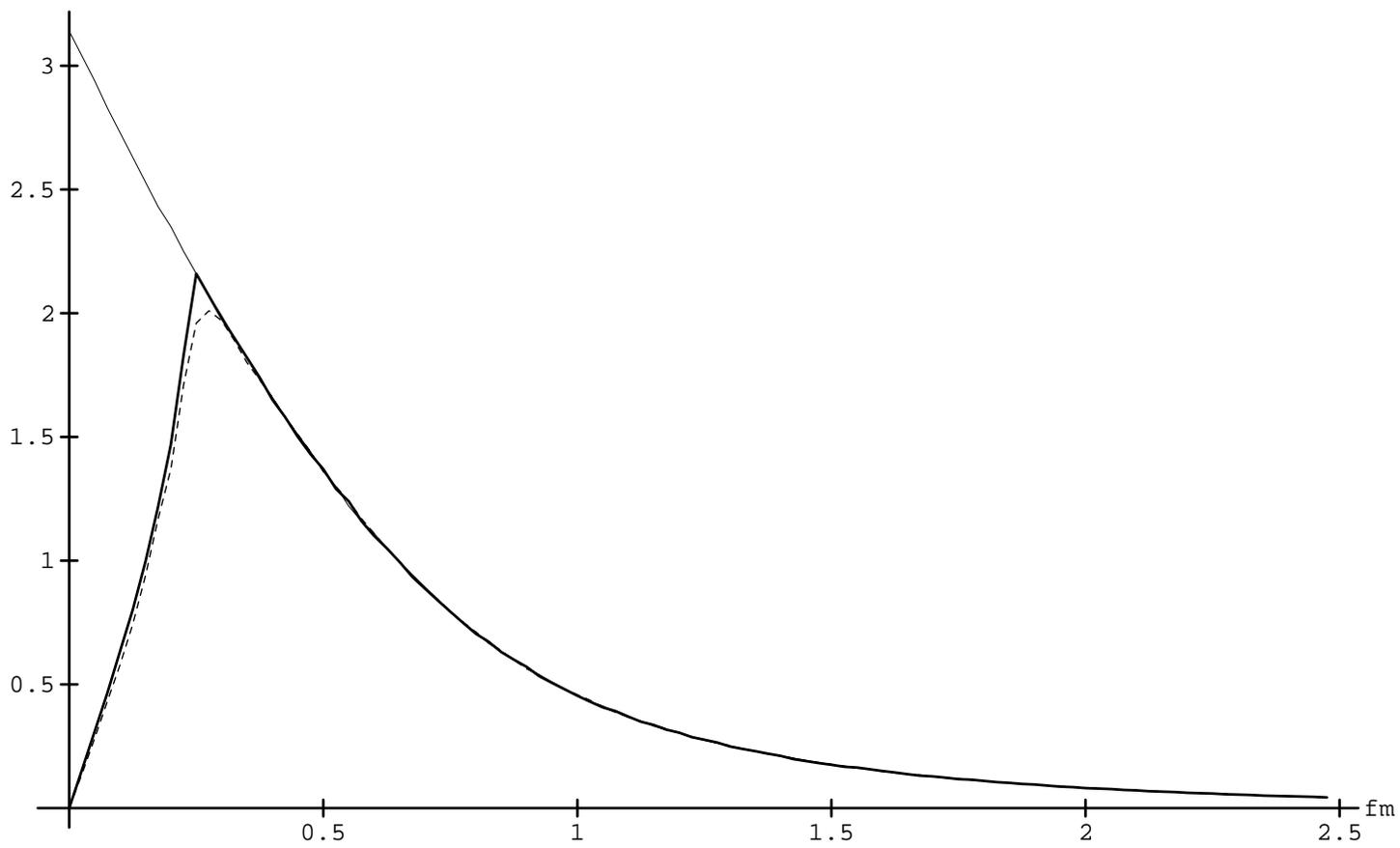
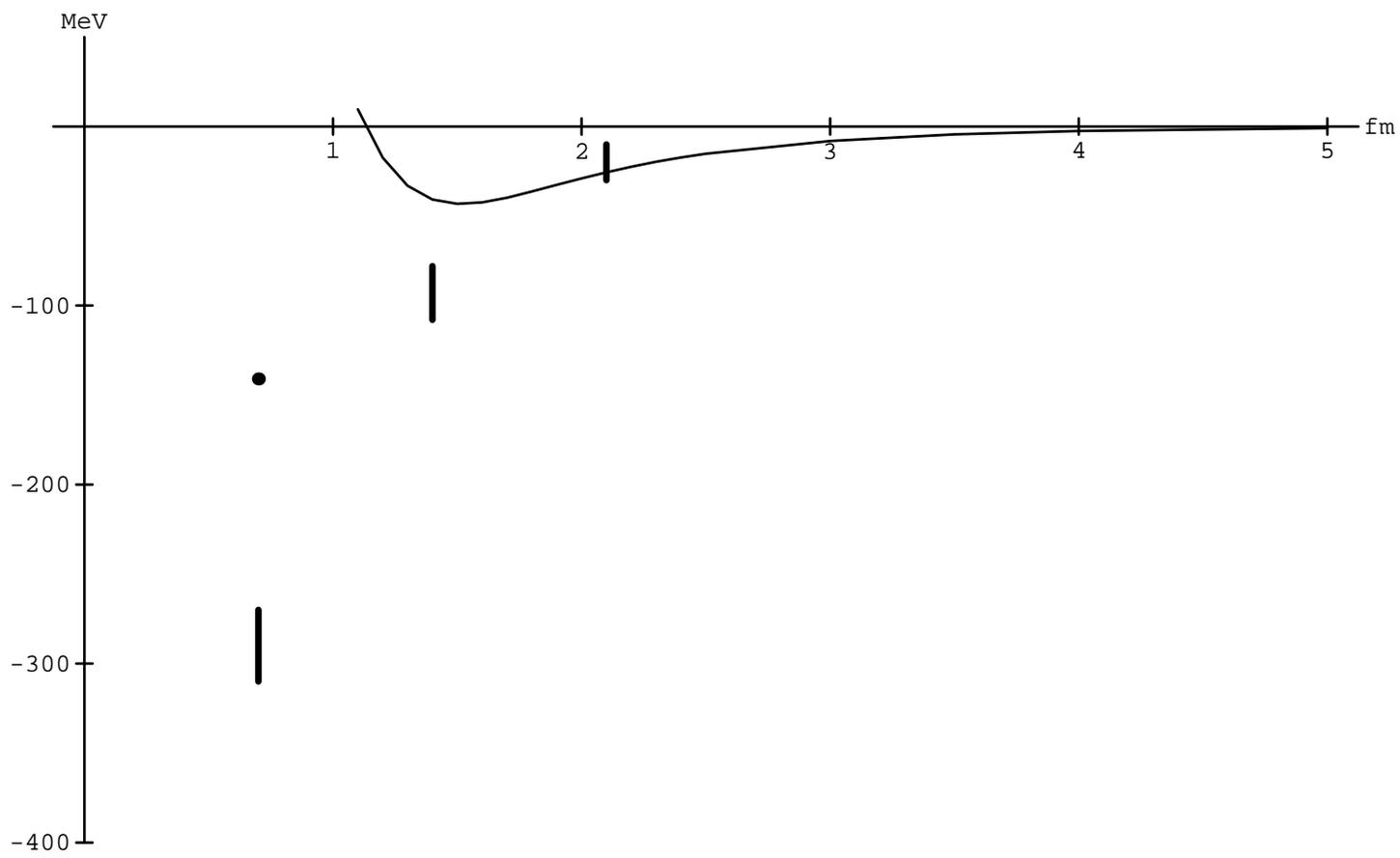
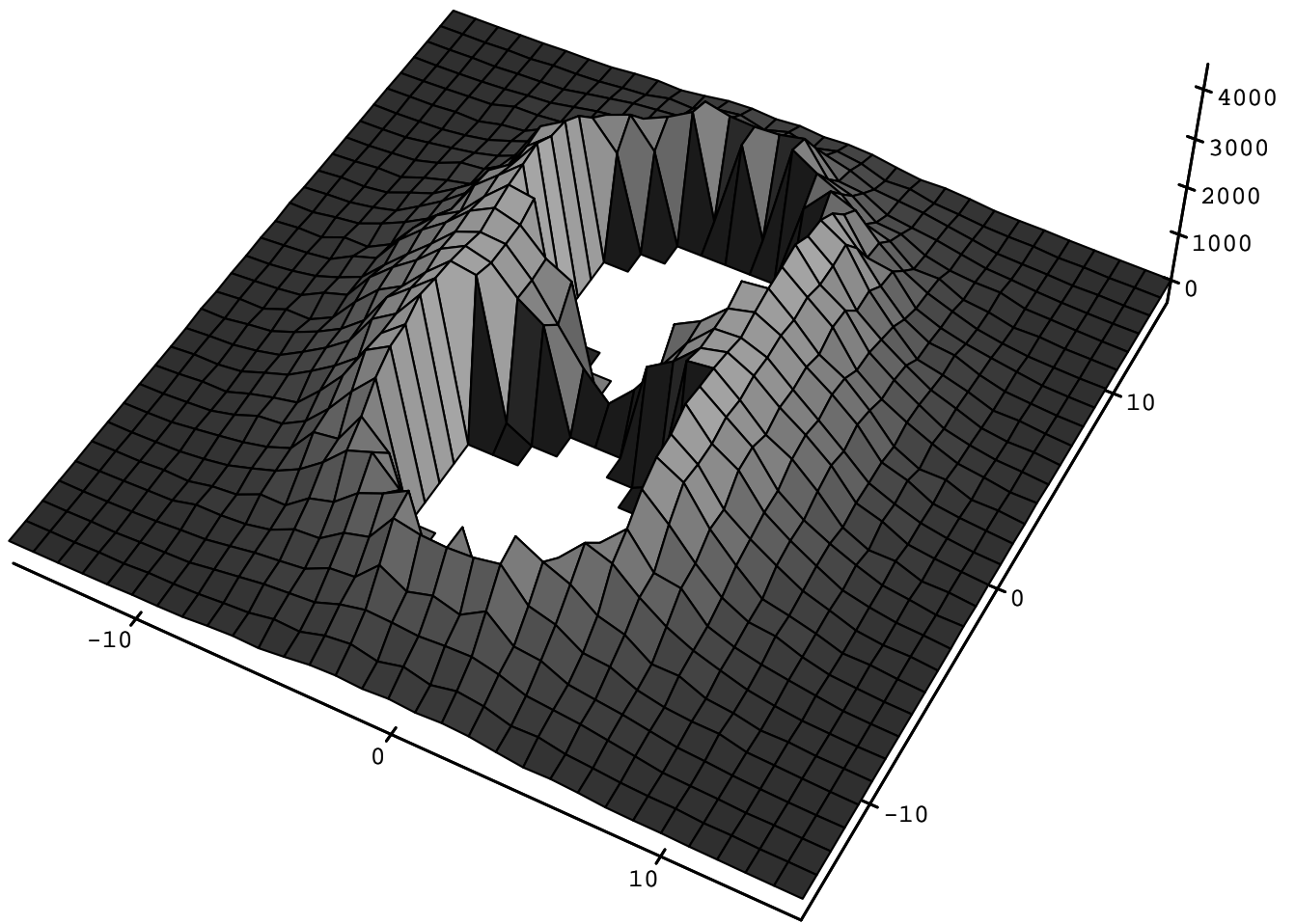
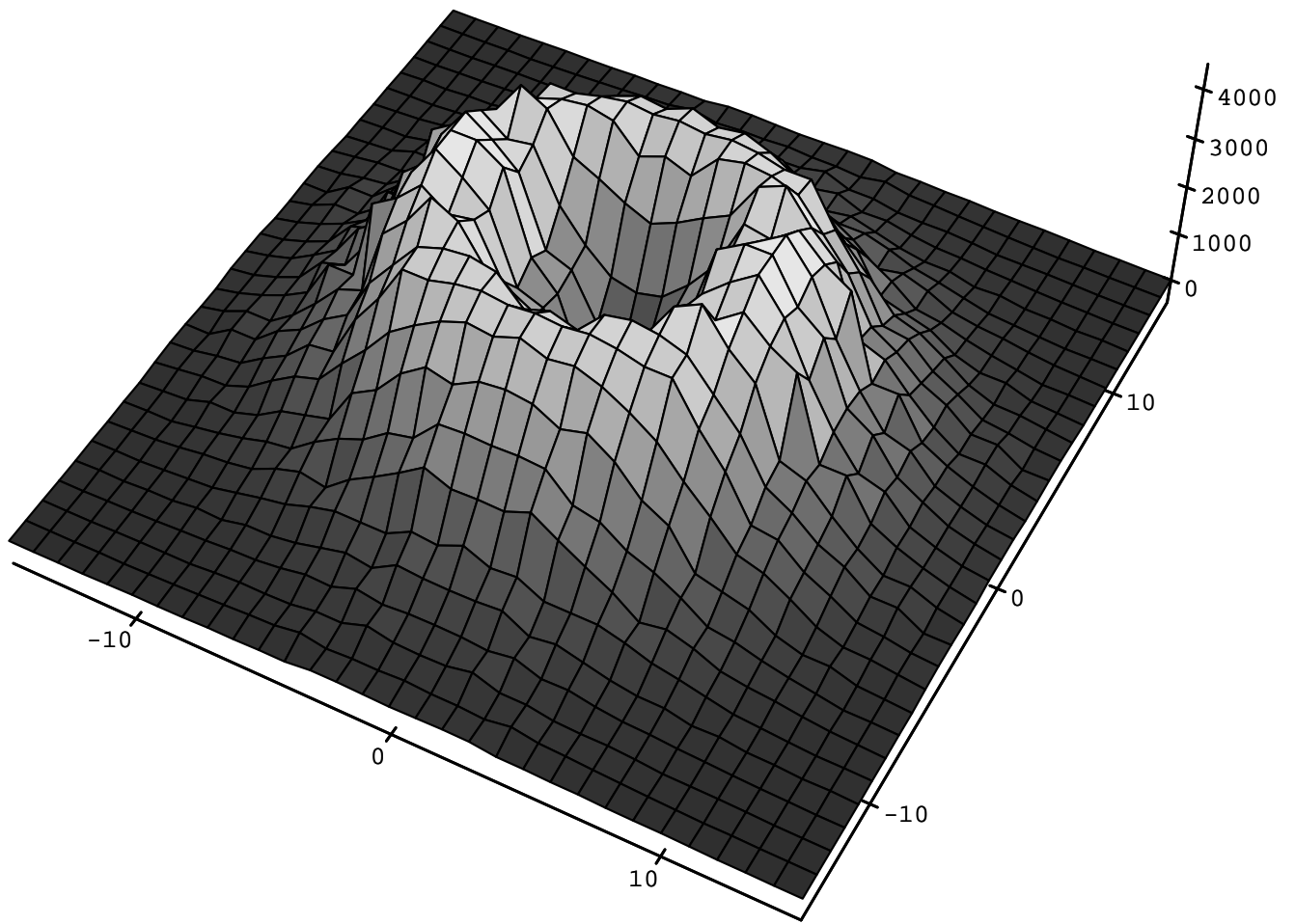
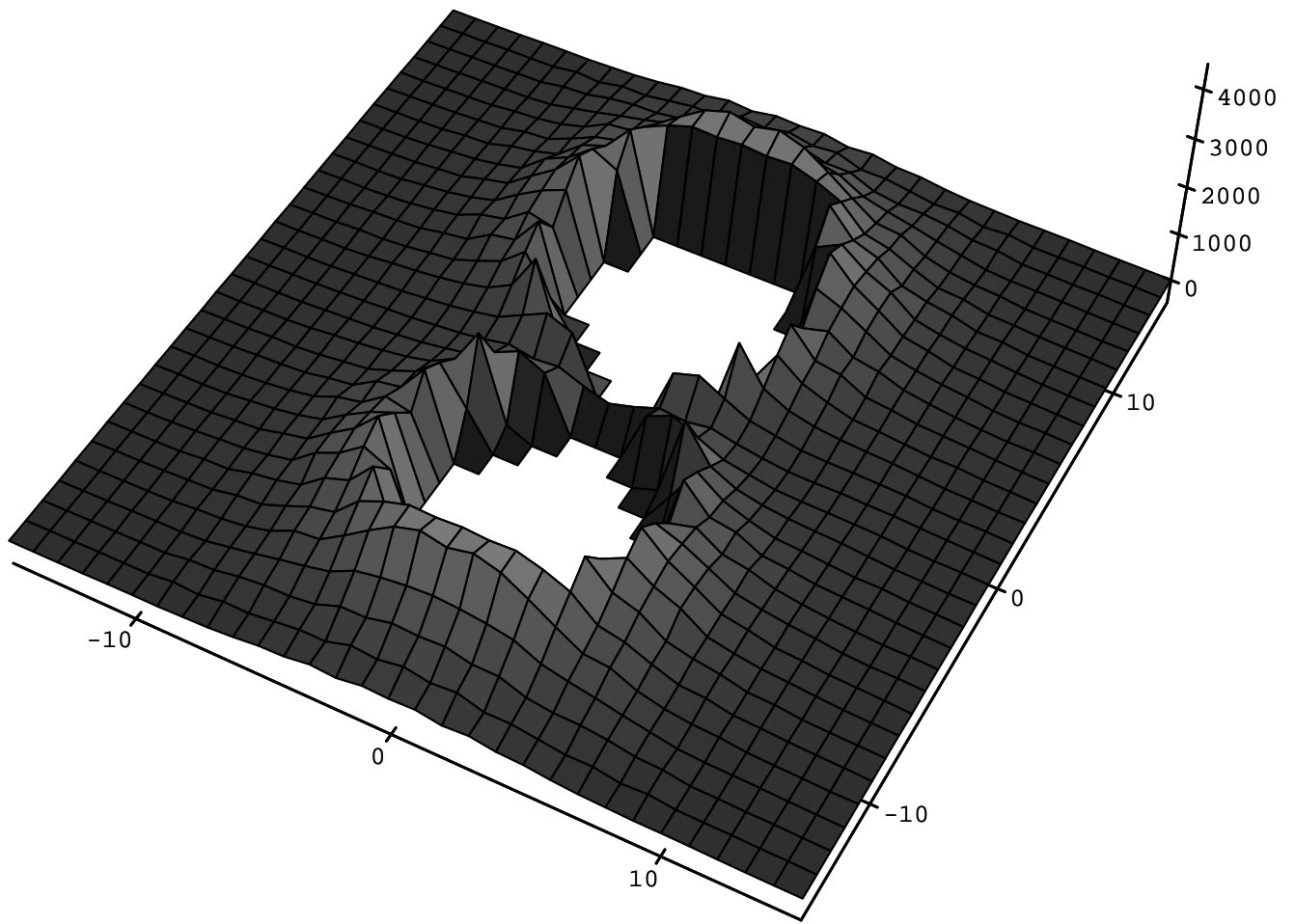
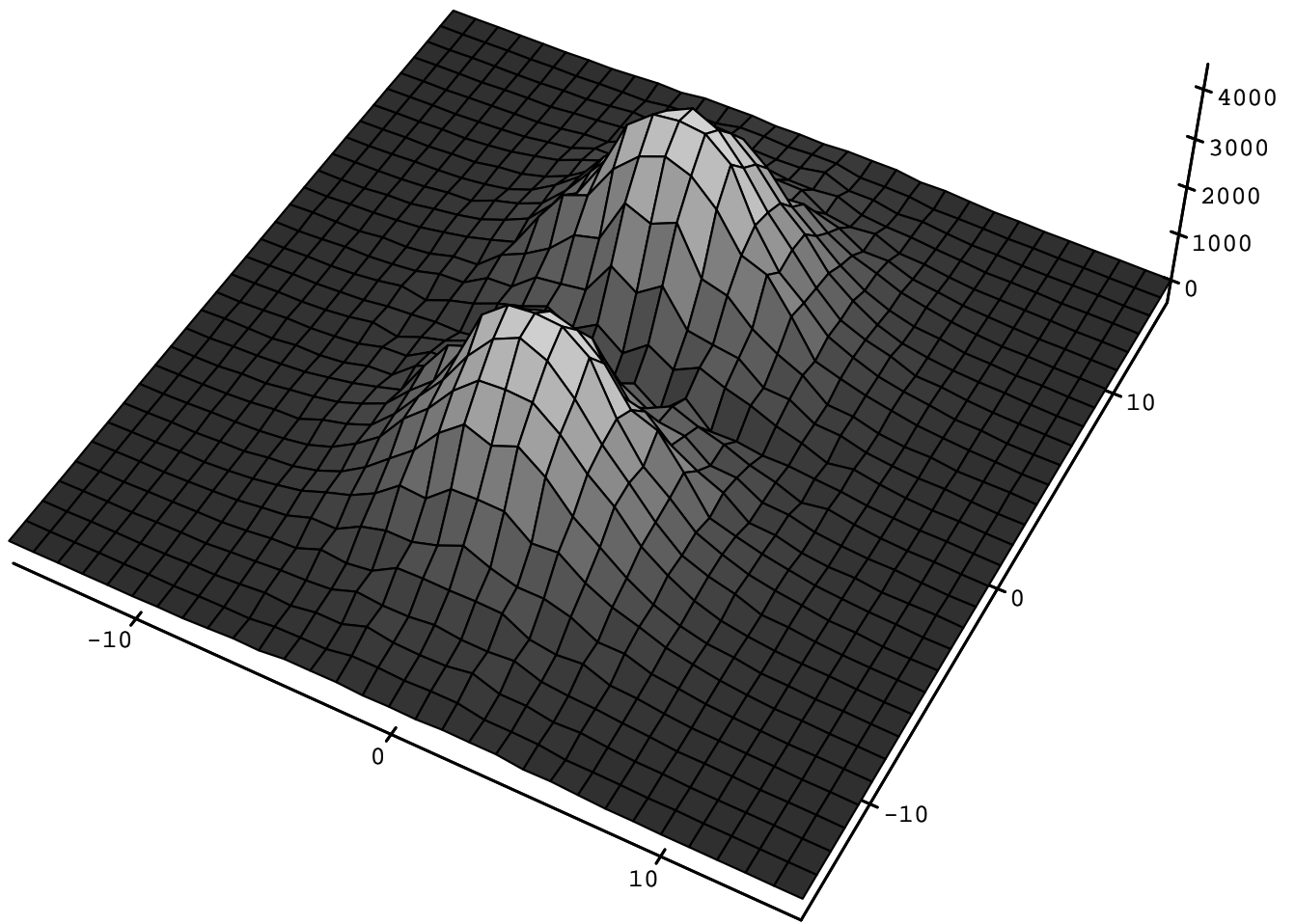
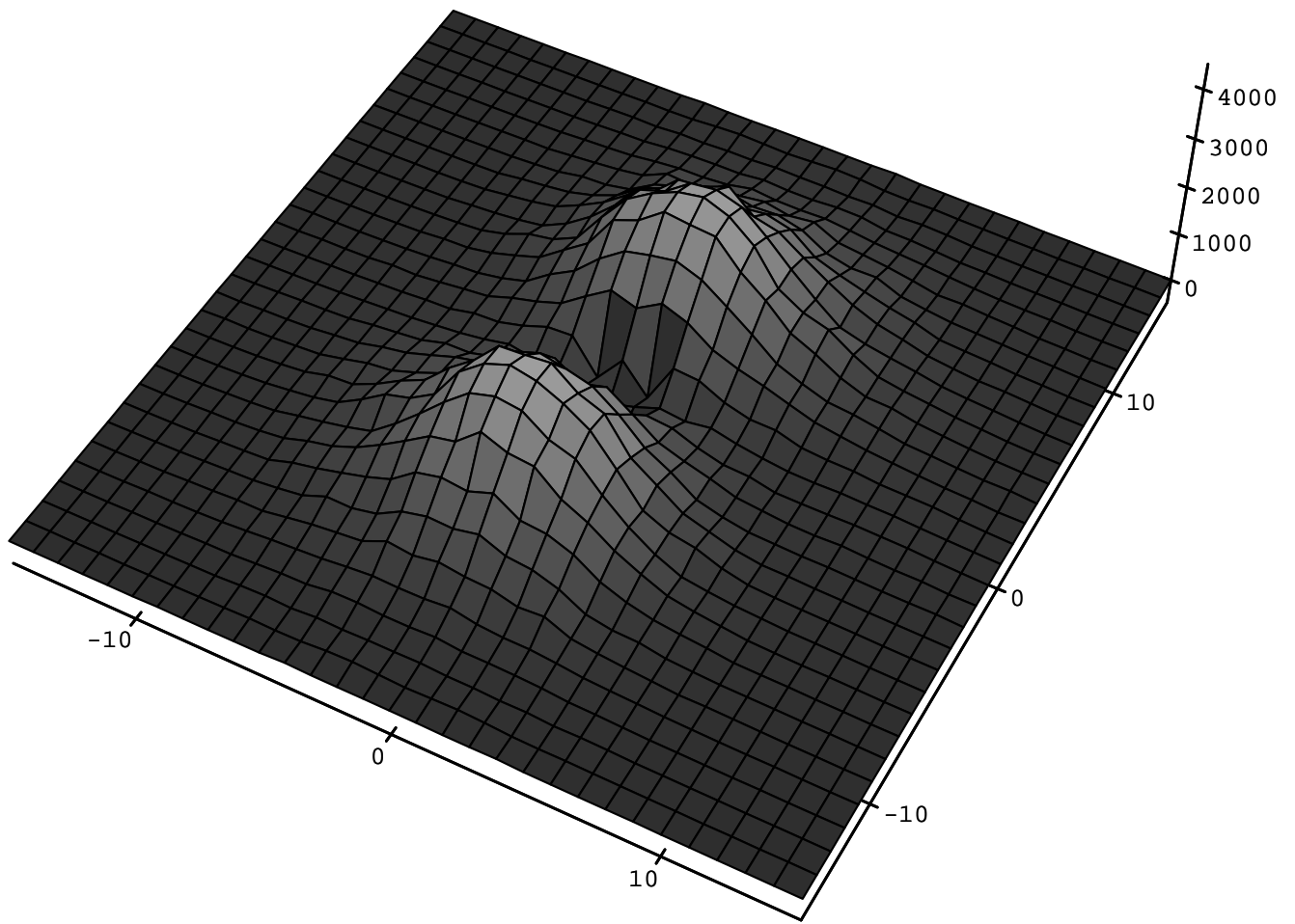
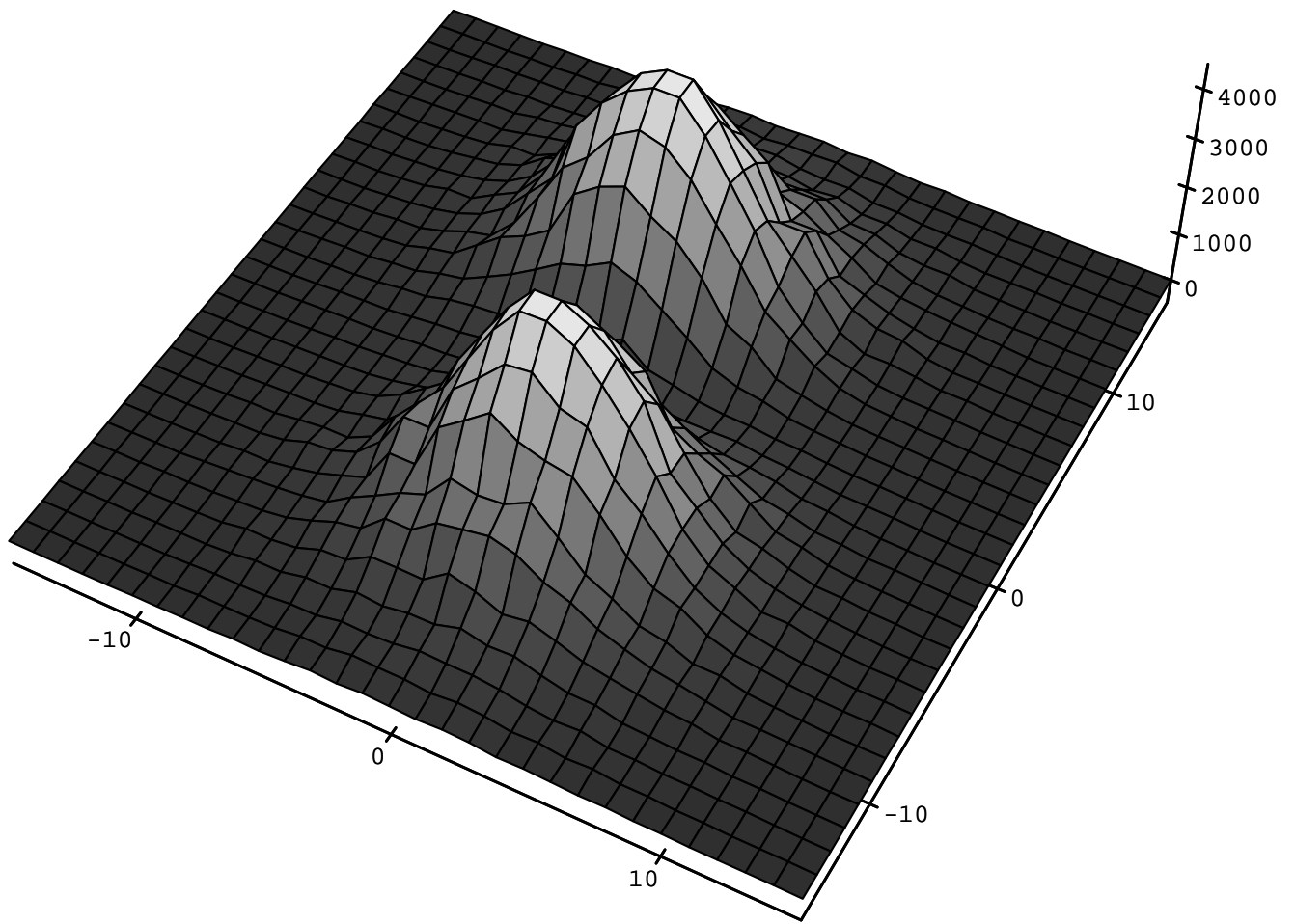
\end{document}